\newcommand{\CLASSINPUTtoptextmargin}{0.8in}
\newcommand{\CLASSINPUTbottomtextmargin}{1in}
\documentclass[conference]{IEEEtran}
\usepackage[utf8]{inputenc}
\usepackage{cite}
\usepackage{amsmath}
\usepackage{graphicx}
\usepackage{algorithm}
\usepackage{algpseudocode}
\usepackage{caption}
\usepackage{enumitem}
\newtheorem{theorem}{Theorem}
\usepackage{hyperref}
\usepackage{caption}
\usepackage{subcaption}
\title{Partitioning and Placement of Deep Neural Networks on Distributed Edge Devices to Maximize Inference Throughput}
\author{
\IEEEauthorblockN{ Arjun Parthasarathy
}
\IEEEauthorblockA{ Crystal Springs Uplands School \\
    \textit{Email: aparthasarathy23@csus.org}}
\and
\IEEEauthorblockN{Bhaskar Krishnamachari}
\IEEEauthorblockA{University of Southern California \\
    \textit{Email: bkrishna@usc.edu}}

}

\date{November 2022}
\begin{document}

\maketitle
\begin{abstract}
    Edge inference has become more widespread, as its diverse applications range from retail to wearable technology. Clusters of networked resource-constrained edge devices are becoming common, yet no system exists to split a DNN across these clusters while maximizing the inference throughput of the system. We present an algorithm which partitions DNNs and distributes them across a set of edge devices with the goal of minimizing the bottleneck latency and therefore maximizing inference throughput. The system scales well to systems of different node memory capacities and numbers of nodes. We find that we can reduce the bottleneck latency by 10x over a random algorithm and 35\% over a greedy joint partitioning-placement algorithm. Furthermore we find empirically that for the set of representative models we tested, the algorithm produces results within 9.2\% of the optimal bottleneck latency.
\end{abstract}

\section{Introduction}
Deep Neural Networks (DNNs) have greatly accelerated machine learning across different disciplines, such as Computer Vision \cite{chai2021deep} and Natural Language Processing \cite{min2021recent}. Edge Inference is becoming an increasingly popular field with multiple facets \cite{wu2021accelerating}, as sensor-driven computation in IoT systems necessitates DNN inference in the field. IoT applications for edge inference range from retail to wearable technology \cite{chen2019deep, biswas2021survey}.

The edge can come in multiple configurations \cite{luo2021resource, sonkoly2021survey}, and there are multiple approaches to facilitate edge inference. For cloud-edge hybrid inference, one such approach is model compression \cite{gholami2021survey}, which deals exclusively with DNN optimization but does not address the system's runtime configuration. In this paper, we focus on clusters of resource-constrained edge devices. These \textit{edge clusters} are becoming increasingly common due to their low-cost and scalability at the edge \cite{premkumar2021survey}. Unlike a cloud data center, the edge brings system resource limitations and communication bottlenecks between devices.

With this in mind, we address the following problem: \textbf{How can we take advantage of multi-device edge clusters to enable high-performance DNN inference while respecting computational resource constraints and taking into account the heterogeneity of communication links?}

\begin{figure}
\centerline{\includegraphics[scale=0.4]{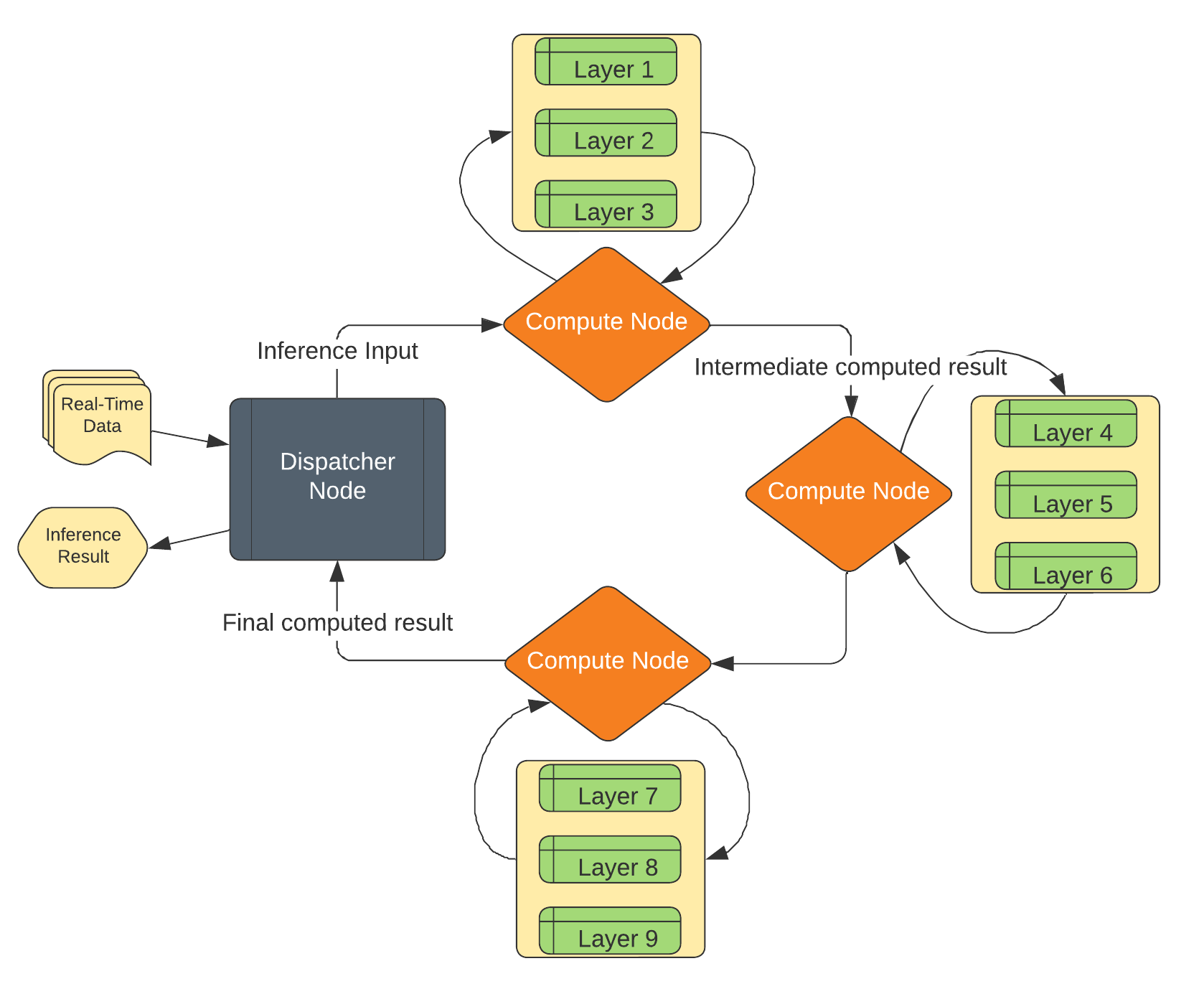}}
\caption{Partitioning and Distributing a Model Across Edge Devices to Create an Inference Pipeline}
\label{fig:partitioning}
\end{figure}

To partition a deep learning model, we first split the model into components that are executed sequentially. Each partition is assigned to a different edge device, and once each node performs inference with its piece of the model, that intermediate inference result is sent to the next node with the corresponding partition in the sequence. This inference pipeline is shown in Figure \ref{fig:partitioning}.

In an edge cluster, although we have a lower computational power in each node, we can take advantage of this inference pipelining to increase system throughput. Since each node can perform inference with its partition individually, prior nodes in the pipeline can send their finished inference results to the subsequent nodes in the pipeline and accept new batches. 

We define the \textit{throughput} metric of a system as the number of inference cycles it can perform per unit time. As we showed in our previous work DEFER \cite{parthasarathy2022defer}, we can achieve higher throughput with distributed edge inference as opposed to inference on a single device because of pipelining. The throughput is defined as the reciprocal of the \textit{bottleneck latency}. For nodes $[k] = \{1, 2, \dots, k\}$, the bottleneck latency $\beta$ is defined as

\begin{equation}
    \begin{array}{c}
        S = \{k \in [k] \mid c_k, \gamma_k\} \\
        \beta = \max_{s \in S}{s} 
    \end{array}
\end{equation}

where $c_k$ is the compute time of the operations on node $k$, and $\gamma_k$ is the communication time between node $k-1$ and $k$. We use ResNet50 \cite{resnet50}, which is a representative model for our use case. On a Raspberry Pi 4, the inference speed was found to be 225 ms \cite{resnetinference}. Next, we found the amount of data transferred between each layer of the model. On average, 10.2 Mbits of data was transferred between layers. Given an average WiFi bandwidth of 6 Mbps for a low-end edge network, this gives us a communication time of $1.7s$. This is 7.5x slower than the compute time. In reality, many models are larger than ResNet50 and will therefore be split across devices, so each device will have less operations to execute. This means that communication time will outweigh compute time as the bottleneck. Therefore, we can simplify the expression for bottleneck latency to the following:

\begin{equation}\label{eq:bottleneck}
    \beta = \max_{k \in [k]}{\gamma_k}
\end{equation}

Since throughput is defined as $\frac{1}{\beta}$, by minimizing the bottleneck latency we maximize inference throughput. Additionally, we assume that all nodes are homogeneous in RAM. If the devices are not the same capacity, then the algorithm will take the smallest memory capacity across all nodes in the cluster, and take that as the capacity of each node. In this paper, we primarily analyze image and text models due to their prevalence on the edge for visual analytics applications \cite{8781894, nayak2021review}. 

Our main contribution in this paper is a novel partitioning and placement algorithm for DNNs across a cluster of edge devices distributed spatially within the same WiFi network. The algorithm finds the candidate partition points, finds the optimal partition sizes to transfer the least amount of data, and finds the arrangement of nodes with the highest bandwidth. Together, these aim to minimize the resulting bottleneck latency according to the throughput metric. We found that our algorithm results in a 10x improvement over a random partitioning/placement algorithm, and a 35\% reduction in bottleneck latency for systems with 50 compute nodes. We  empirically observe an average approximation ratio of 1.092 for the bottleneck latency (i.e. it is 9.2\% more than the optimal bottleneck latency, on average).

\section{Related Work}

Early works on the topic of partitioning DNN models divided them into head and tail models with the former distilled to enable running on a resource-constrained device and reduce data transfer~\cite{matsubara2019distilled}. Some prior works on DNN edge inference mathematically perform DNN model slicing by layer \cite{zhang2020dynamic, zhang2022teeslice}, after calculating layer impact during the training stage; these do not account for communication demands on the edge. Others abstract model layers into certain ``execution units," \cite{li2021slicing, cho2021dnn} which they then choose to slice based on certain resource requirements. Li \emph{et al.} \cite{li2018edge} regressively predict a layer's latency demand and optimize communication bandwidth accordingly. DeeperThings \cite{stahl2021deeperthings} performs layer fusion on CNNs to optimize data transfer. These works are optimized for a hybrid edge-cloud pipeline and do not address the demands of a cluster of edge devices. Couper \cite{hsu2019couper} uses a similar partitioning scheme to minimize inter-partition data transfer, but does not address the communication bottleneck associated with an edge cluster. Hu \emph{et. al} \cite{hu2020fast} optimize the partitioning of a CNN onto a set of devices by taking compute time as a bottleneck, while employing compression to deal with communication constraints, and do not consider placement. Our paper builds on and differentiates itself from these works by addressing the bandwidth limitation of an edge cluster, and aims to maximize inter-node bandwidth during the placement stage to minimize bottleneck latency.

\section{Partitioning and Placement Algorithm}
We are given two graphs:

\begin{enumerate}
    \item An unweighted DAG $G_m$ representing the computation graph of a DNN, where each vertex represents a layer in the model. This DAG can be found using common ML libraries such as Tensorflow \cite{tensorflow} and Keras \cite{keras}.
    \item A weighted complete graph $G_c$ representing the communication graph of a cluster of homogeneous physical compute nodes, where each vertex represents a physical compute node and each edge represents the bandwidth between those nodes. The graph is complete because we assume that these edge devices will communicate over the same WiFi network.
\end{enumerate}

Our goal is to optimally partition the model and place these partitions on a set of edge devices. We do so as follows.

\subsection{Converting a Complex DAG to a Linear DAG}

\begin{figure}
\centerline{\includegraphics[scale=0.3]{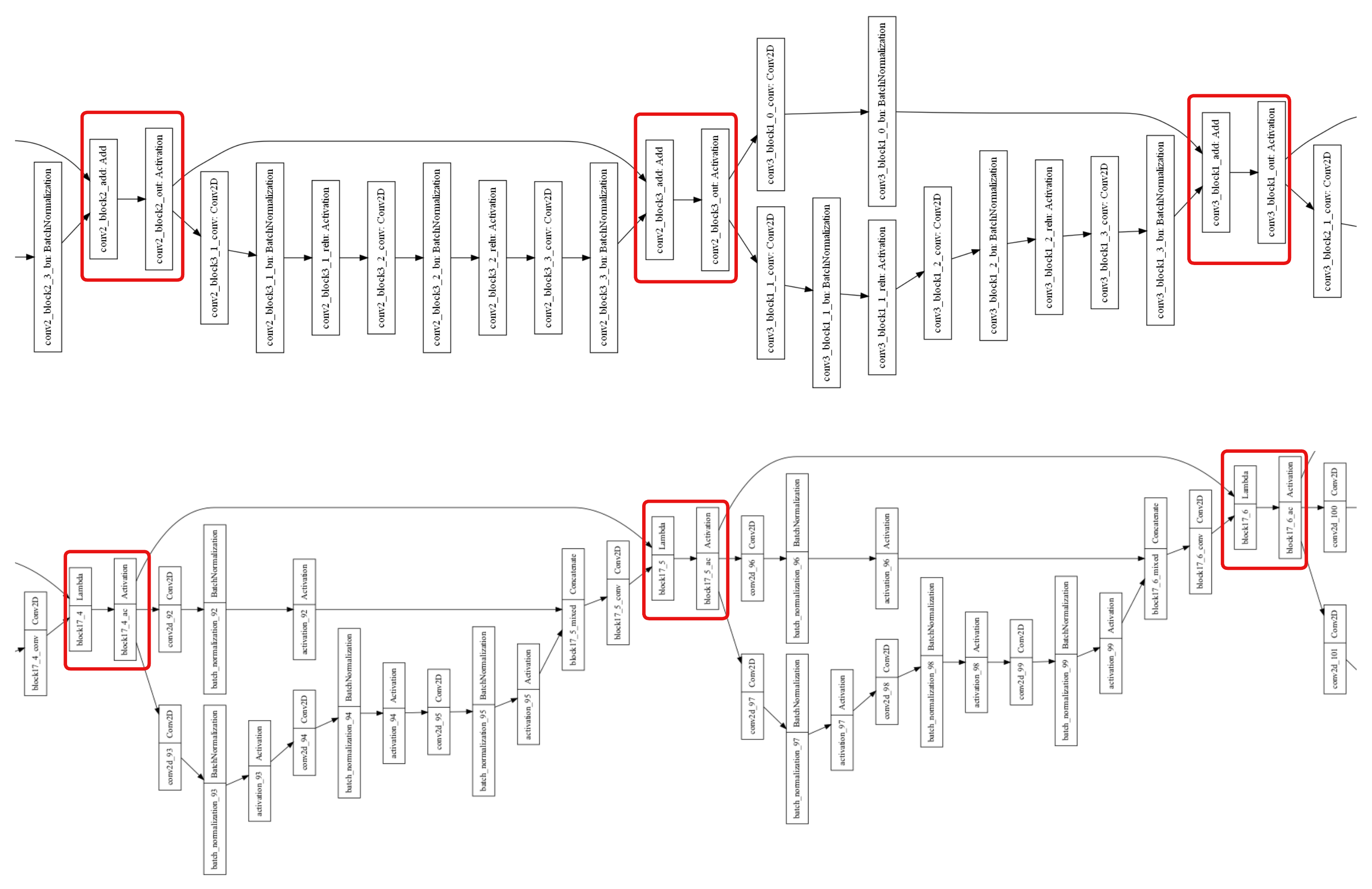}}
\caption{Partition points for ResNet50 and InceptionResNetV2 models}
\label{fig:examplepartpts}
\end{figure}

First, we need to distill $G_m$ into a linear DAG. The vertices where it is possible to partition the model are called ``candidate partition points." We illustrate this in Figure \ref{fig:examplepartpts}.

For $v \in V$, edges $e \in E$ and source vertex $s$ of $G_m$, find the longest path from $s$ to $v$. This can be done by topologically sorting the DAG and for each vertex in the resulting list, relaxing each neighbor of that vertex. We call the length of this longest path the \textit{topological depth} of that vertex in the graph. Let $LP(v)$ denote the length of longest path from $s$ to $v$.

To verify that all paths from vertex $v_{prev}$ go through vertex $v$, use a modified DFS by recursing on the incident edges of each vertex. If we encounter a vertex with a greater topological depth than $v$, return false. If we reach vertex $v$, return true. Let $AP(v_{prev}, v)$ denote the result of this algorithm.

Given the previously found candidate partition point $p_{k-1}$ and the current vertex $u$, the next candidate partition point $p_k = u$ iff:

\begin{enumerate}
    \item $LP(u) \neq LP(v) \forall v \in \{V - u\}$
    \item $AP(p_{k-1}, u) = \text{true}$
\end{enumerate}

with $p_0 = s$.

The time complexity of LP is $O(V+E)$. AP runs in polynomial time by returning upon reaching a vertex with a greater topological depth. Therefore, this algorithm runs in polynomial time.

Figure \ref{fig:examplepartpts} shows the candidate partition points at certain sections of the DAG of ResNet50 \cite{resnet50} and InceptionResNetV2 \cite{inceptionresnetv2}. Each rectangle represents a model layer in the DAG.

\begin{figure}
\centerline{\includegraphics[scale=0.4]{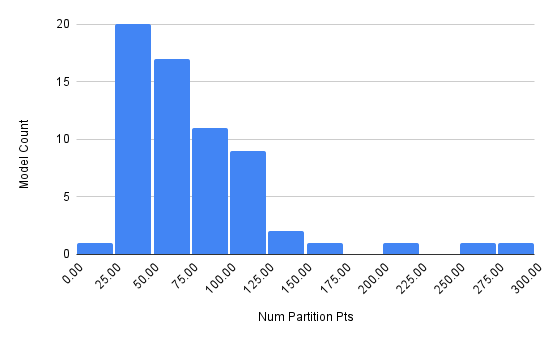}}
\caption{Histogram of Number of Candidate Partition Points}
\label{fig:numpartpts}
\end{figure}

We then calculated the number of partition points for the set of Keras pretrained models. As shown in Figure \ref{fig:numpartpts}, almost all the models have at least 25 candidate partition points. There are some model architectures, like NASNet \cite{nasnet}, which do not allow partitioning under our scheme.

\begin{figure}
\centerline{\includegraphics[scale=0.3]{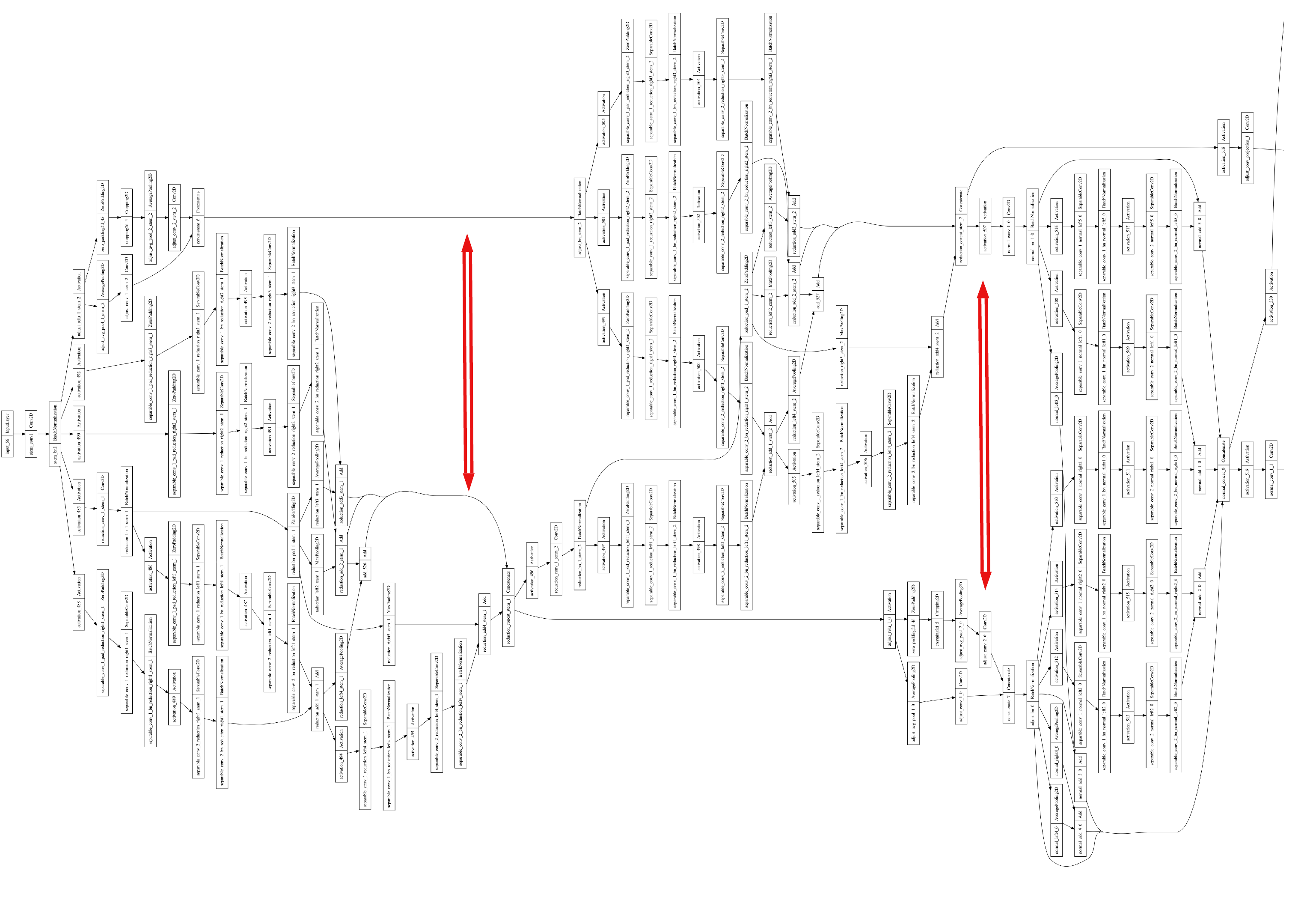}}
\caption{Portion of NASNet's layer DAG}
\label{fig:nasnet}
\end{figure}

As shown in Figure \ref{fig:nasnet}, NASNet cannot be partitioned because there is no single point that splits the model into a distinct execution unit that does not have any dependencies to a previous or subsequent layer. If we run our LP algorithm, we find that there is no single layer that has distinct topological depth from other layers. We found that 64 of the 66 (97 \%) pretrained Keras models \cite{kerasmodels} could be partitioned under our scheme, and only the NASNet variants could not.

\subsection{Optimal model partitioning and placement}

Our goal is to maximize throughput of the system. As previously discussed, this means we need to minimize the bottleneck latency. Latency is defined as $\frac{\text{data}}{\text{bandwidth}}$. Given a tuple of partition points $P_{opt}$, their transfer sizes $T$, and a set of bandwidths $B$ between compute nodes, the latency between each set of compute node is defined as 

\begin{equation}\label{eq:latency}
    \gamma_k = \frac{T_{opt, k}}{B_k} \forall 0 \leq k < \lvert P_{opt} \rvert
\end{equation}

The bottleneck latency for the system is then given by Equation \ref{eq:bottleneck}. For the purposes of explanation, we separate the problems of optimizing the partitions (thereby optimizing transfer size) and optimizing placement (thereby optimizing bandwidth between nodes). We show empirically that this results in the the smallest bottleneck latency. In Section \ref{section:results}, we compare this formulation to an algorithm that tries to jointly optimize transfer size and bandwidth.

\subsubsection{Finding optimal partitions}
Our heuristic for finding optimal partitions is the ``transfer size" of the partition; i.e how much data will be transferred from that partition to the next. Given the tuple of candidate partition points $P = (p_0, p_1, \dots, p_k)$, we now need to find a set of model partitions which minimizes the sum of transfer sizes. Assuming a batch size of 1, the transfer size $t_k$ of candidate partition point $p_k$ is defined as

\begin{equation}
    t_k = \frac{\eta}{\lambda}
\end{equation}

Given a floating point array representing the output of the model layer $l_k$ (which is the same layer given by the candidate partition point $p_k$), $\eta$ represents the size of that array.

$\lambda \approx 1.44 * 2.1$ represents the total compression ratio given by multiplying the average ZFP compression ratio \cite{zfp} by the average LZ4 compression ratio \cite{lz4}.

To better illustrate our algorithm, we classify the transfer size $t_k$ into 3 transfer size classes (``low'', ``medium'', or ``high'') based on the distribution of the transfer sizes.

\begin{equation}
\label{eq:classes}
\begin{array}{lr}
C = \{L, M, H\} & t_k \subseteq C
\end{array}
\end{equation}

The optimal set of partitions is the scheme which minimizes the sum of the transfer sizes of said partitions.
\\
Let $G_p$ represent a DAG, where each vertex is represented by a possible partition. The vertices are defined as follows: 

\begin{equation}
\begin{array}{l}
    p_i, p_{i+1}, \dots, p_j\}) < \kappa \mid \{p_i, p_{i+1}, \dots, p_j\}\} \\
    \forall {0 \leq i < \lvert P_{opt} \rvert, 0 \leq j < \lvert P_{opt} \rvert - i}
\end{array}
\end{equation}

The set of vertices represents every possible contiguous subarray of candidate partition points, where $\omega(P)$ finds whether the memory use of partition $P$ is within the memory capacity $\kappa$ of the compute node. We quantize the models using TFLite \cite{tflite} quantization to reduce their memory footprint. However, when calculating the memory footprint of a partition, we do not consider this quantization. This means that we are conservative on partition size and in turn provide extra space on each device for the memory overhead from containerization. Each partition is a set of layers that fall between the partition points $p_i$ and $p_j$.

The set of edges is defined as follows:

\begin{equation}
    E = \{(u, v) \in V, \rho(u_{\lvert u \rvert - 1}) = \rho(v_0 - 1) \mid (u, v)\}
\end{equation}

The function $\rho(\upsilon)$ finds the index of element $\upsilon$ in $P_{opt}$. There is an edge between vertices if the last partition point of $u$'s partition is adjacent in $P$ to the first partition point of $v$'s partition. For example, if $u = [1, 2]$, $v = [3, 4]$, and $P = (1, 2, 3, 4)$, then $(u, v)$ is an edge. Each edge has a weight $w(u, v)$ which corresponds to its transfer size class.

\begin{figure}
\centerline{\includegraphics[scale=0.5]{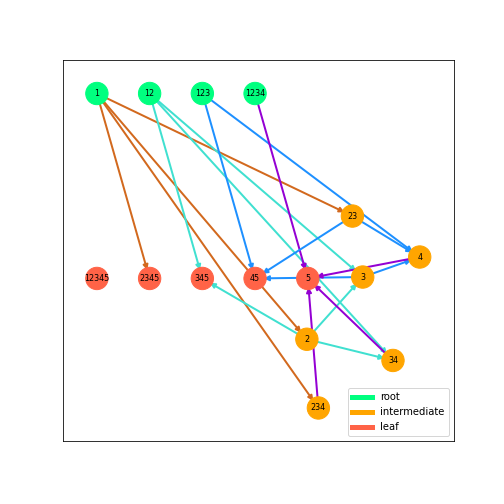}}
\caption{Example partition graph, where the partition points are $P = \{1, 2, 3, 4, 5\}$}
\label{fig:partitiongraph}
\end{figure}

Figure \ref{fig:partitiongraph} shows an example partition graph, where edges that are the same color will have the same weight. In the figure, ``root'' vertices have in-degree 0, ``leaf'' vertices have out-degree 0, and ``intermediate'' vertices have neither.

\begin{algorithm}
\caption{Optimal Partitioning}\label{alg:partitioning}
\begin{algorithmic}
\State{\textbf{//} Map to store memoized paths}
\State{$pathFrom \gets \Call{NEW-MAP}$} \label{line:memoize}
\Procedure{MIN-COST-PATH}{$G$, $v$}
    \If {$v.children = \emptyset$}
        \State \Return{$v, 0$}
    \EndIf
    \State{$partitionLastLayer \gets v[v.length - 1]$}
    \If{$partitionLastLayer \notin pathFrom$}
        \State{$paths \gets []$}
        \For {$c \in v.children$}  
            \State{$path, cost \gets \Call{MIN-COST-PATH}{G, c}$}
            \State{$paths \gets \Call{APPEND}{paths, (path, cost)}$}
        \EndFor
        \State{$pathFrom[partitionLastLayer] = \Call{MIN}{paths}$}
    \EndIf
    \\
    \State{$best \gets pathFrom[partitionLastLayer]$}
    \State{$minPath, minCost \gets best$}
    \State{$chosenNode \gets minPath[0]$}
    \State{\textbf{//} Path starting at v and going to a leaf}
    \State{$newPath \gets \Call{APPEND}{[v], ...minPath}$}
    \State{$newCost \gets minCost + w(v, chosenNode)$}
    \State \Return{$newPath, newCost$}
\EndProcedure
\\
\Procedure{PARTITION}{G}
\State{$roots \gets \Call{GET-ROOT-VERTICES}{G}$}
\For {$r \in roots$}
    \State{$path, cost \gets \Call{MIN-COST-PATH}{G, r}$}
    \State{$paths \gets \Call{APPEND}{paths, (path, cost)}$}
\EndFor
\State{$minPath, minCost \gets \Call{MIN}{paths}$}
\State \Return minPath
\EndProcedure
\\
\State{$\Theta \gets \Call{PARTITION}{G_p}$}
\end{algorithmic}
\end{algorithm}

Algorithm \ref{alg:partitioning} finds the shortest path in the graph from a root to a leaf. Since edges which bridge the same candidate partition points (and have the same color as shown in Figure \ref{fig:partitiongraph}) will have the same subsequent paths, we can memoize the shortest path. On line \ref{line:memoize}, we store a map on which tells us for each candidate partition point what the shortest path is from that point. Using memoization, Algorithm \ref{alg:partitioning} takes $O(N)$ to find the shortest path, but $O(N^2)$ to construct the partition graph. Therefore the runtime of Algorithm \ref{alg:partitioning} is $O(N^2)$, where N is the number of nodes.

Let $\Theta$ represent the set of chosen partitions. For each subarray in $\Theta$, we take the last element of the subarray, add that to the list of partition points $Q$, and add its corresponding transfer size to the list $S$. The resulting list is then sorted based on the topological depth of each partition point, so that the partitions are executed in the order they appear in the model.

\subsubsection{Finding optimal model placement}
\label{section:modelplacement}
With the set of optimal partitions $Q$ and their corresponding transfer sizes $S$, we now need to ``match" them to the vertices of $G_c$. We know from Equation \ref{eq:classes} that $S \subseteq C$. Let $c(e)$ return the bandwidth class of a given edge of $G_c$. We use the following threshold function to classify each edge:

\begin{equation}
    \tau(X, t) = \left\{
    \begin{array}{lr}
        c(e) = C_{\arg X - 1}, & \text{if } e < t\\
        c(e) = X, & \text{if } e \geq t
    \end{array}
\right\}{\forall e \in E_c}
\end{equation}
If the edge is greater than or equal to the threshold, it will be classified as class $X$, otherwise it will be classified as the class in $C$ right below $X$. In order for our algorithm to work, we set the number of transfer size classes equal to the number of bandwidth classes.

\begin{figure}
\centerline{\includegraphics[scale=0.5]{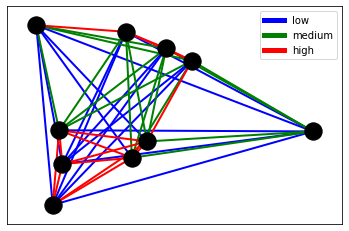}}
\caption{Example communication graph with different bandwidth classes}
\label{fig:commgraph}
\end{figure}

Figure \ref{fig:commgraph} shows an example communication graph, where the nodes are colored black and different bandwidth classes of edges are shown.

Given the array of transfer sizes $S$ and array of communication graph edges $E_c$, the lower bound on bottleneck latency we can achieve is given by Theorem \ref{thm:latencybound}.

\begin{theorem}\label{thm:latencybound}
    The lowest bottleneck latency we can achieve is:
    \begin{equation}\label{eq:minlatency}
        \min(\beta) = \frac{\max{S}}{\max{E_c}}
    \end{equation}
    Therefore, if we achieve $\min(\beta)$, then we have found the optimal minimum bottleneck latency.
\end{theorem}

We prove Theorem \ref{thm:latencybound} as follows:

Given the highest transfer size ($\max S$), then it must be matched with the highest bandwidth ($\max E_c$) to have the lowest bottleneck latency. There are two cases in which the system would have another bottleneck latency:

\begin{enumerate}
    \item 
        \begin{equation}\label{eq:highmismatch}
            \beta = \frac{\max{S}}{e}\forall e \in E_c - max(E_c)
        \end{equation}
    \item 
        \begin{equation}\label{eq:higherlatency}
        \begin{array}{l}
            \beta = \frac{s}{e} \\
            \forall \{s \in S - max(S), \\ e \in E_c - max(E_c) \mid \beta \geq \frac{\max{S}}{\max{E_c}}\}
        \end{array}
        \end{equation}
\end{enumerate}

In Equation \ref{eq:highmismatch}, the latency of the system would be higher than \ref{eq:minlatency}, since the transfer size is being matched with a lower bandwidth edge. In Equation \ref{eq:higherlatency}, some other transfer size $s$ and bandwidth $e$ may result in a higher bottleneck latency, in which case Equation \ref{eq:minlatency} still holds. Therefore, Theorem \ref{thm:latencybound} holds.

We run tests in Section \ref{section:results} to see how often we get this optimal solution. Algorithm \ref{alg:subgraphkpath} performs the matching between $S$ and $G_c$ to try to reach the optimal latency as outlined above. Let $N$ represent the array of nodes that we choose from $G_c$, with length $\lvert S \rvert$.

\begin{algorithm}
\caption{Finding K-Paths}\label{alg:subgraphkpath}
\begin{algorithmic}
\Procedure{SUBGRAPH-K-PATH}{$X$, $k$, $s$, $u$}
\State{\textbf{//} Sort by weight in descending order}
\State{$edgeList \gets \Call{SORT}{G_c, \{e \in E_c \mid w(e)\}, reverse}$} \label{line:sortedgelist}
\State{$low \gets 0$}
\State{$high \gets edgeList.length$}
\State{$bestPath \gets []$}
\While{$low < high$}
    \State{$median \gets \frac{(low + high)}{2}$}
    \State{$\tau(X, edgeList[median])$}
    \State{$\tau(X, threshold)$}
    \State{\textbf{//} Induced subgraph of $G_c$ w/ class $X$ edges}
    \State{$G^X_c \gets \{E^X_c = \{e \in E_c \land c(e) = X \mid e\} \mid V^X_c, E^X_c\}$}
    \State{$result \gets \Call{K-PATH}{G^X_c, k, s, u}$}
    \If{$result = \text{FALSE}$}
        \State{$low \gets median + 1$}
    \Else
        \State{$high \gets median$}
        \State{$bestPath \gets result$}
    \EndIf
\EndWhile
\For{$N \in bestPath$}
    \State{$\Call{DEL}{G_c, N}$}
\EndFor
\EndProcedure
\end{algorithmic}
\end{algorithm}

In algorithm \ref{alg:subgraphkpath}, we use the color-coding k-path algorithm \cite{alon1995color}, which finds a path of length $k$ (where $k$ is the number of vertices) in $G^X_c$ if a $k$-path exists and does so in polynomial time if $k < \log(|V^X|)$. As we show in Section \ref{section:results}, this is possible using node memory capacities which reflect real-world devices. We use a binary search to find the maximum threshold for which a $k$-path exists. On line \ref{line:sortedgelist}, we sort in descending order so that we can find the maximum viable edge-weight threshold with a binary search. As $N$ starts to be filled in, the $k$-paths have to be found between certain nodes in order for $N$ to be a contiguous path of nodes. We modify the $k$-path algorithm to start at $s$ and stop once it reaches $u$. We make the algorithm more efficient by stopping a particular iteration if we reach $u$ before we have a path of length $k$. If $s$ is $null$, find any find any $k$-path that ends at $u$. Similarly, if $u$ is $null$, find any $k$-path that starts at $s$. 

\begin{algorithm}
\caption{K-Path Matching}\label{alg:kpathmatching}
\begin{algorithmic}[1]
\Procedure{K-PATH-MATCHING}{S, C}
\For{$X \in C$} \label{line:xmatch-begin}
    \State{$x\_paths \gets \Call{FIND-SUBARRAYS}{S, X}$} \label{line:findsubs}
    \State{$x\_paths \gets \Call{SORT}{x\_paths, \{p \in x\_paths \mid p.length\}}$}
    \For{$i \gets 0 \textbf{ to } x\_paths.length$}
       \State{$startIdx \gets \Call{INDEX-OF}{x\_paths[i][0], S}$} 
       \State{$x\_len \gets x\_paths.length$}
       \State{$startV \gets N[startIdx]$} \label{line:startV}
       \State{$endV \gets N[startIdx + x\_paths[i].length + 1]$} \label{line:endV}
       \State{$path \gets \Call{SUBGRAPH-K-PATH}{X, x\_len + 1, startV, endV}$}
       \State{$N[startIdx:startIdx + path.length] \gets path$} \label{line:dots2}
    \EndFor
\EndFor \label{line:xmatch-end}
\EndProcedure
\end{algorithmic}
\end{algorithm}

\bigbreak

Algorithm \ref{alg:kpathmatching} performs the $k$-path matching of partitions onto vertices of $G_c$. 

FIND-SUBARRAYS() on line\ref{line:findsubs} returns a list of subarrays of a certain class, by iterating over the list of transfer sizes $S$.

Lines \ref{line:xmatch-begin}-\ref{line:xmatch-end} match paths for all bandwidth classes, starting with class $H$. On line \ref{line:startV}, $startV$ represents the vertex before the current $k$ path. If this is equal to $null$, meaning that the algorithm hasn't reached the iteration of finding the previous $k$-path, then SUBGRAPH-K-PATH will find a path that starts at any vertex. Similarly, on line \ref{line:endV} $endV$ represents the vertex of the current $k$-path. If this is equal to $null$, meaning that the algorithm hasn't reached the iteration of finding that subsequent $k$-path, then SUBGRAPH-K-PATH will find a path that ends at any vertex.

By starting with the longest $H$-subarrays and working to the shortest $L$ subarrays, we are greedily finding the best bandwidth paths to match with the highest transfer size terms of $S$. We continue this process until we have found $k$-path matchings for all subarrays of $S$. In some cases, a high number of bandwidth classes will prevent the algorithm from returning a result, because it has very few edges to choose from during each iteration of the matching. In this case, we can re-run the algorithm with fewer bandwidth classes.

\section{Evaluation Methodology}
\label{section:methods}
We simulated a set of randomly placed edge devices using a random complete graph. For each evaluation, we created a random complete graph by drawing the positions of the nodes from a uniform distribution with the range $x, y \in (-150, -1) \cup (1, 150)$ to simulate a WiFi router with a range of 150m. Between each set of nodes, we calculated the edge weight using Shannon's capacity equation, assuming that the SNR decreases proportionally to the inverse square of the device's distance from the router. 

\begin{equation}
\begin{array}{l}\label{eq:bwdistance}
r(x, y)=\log_{2}\left(1+\frac{a}{(\sqrt{x^2+y^2})^2}\right)=\log_{2}\left(1+\frac{a}{x^2+y^2}\right) \\
x, y \in (-B, -1) \cup (1, B)
\end{array}
\end{equation}

In Equation \ref{eq:bwdistance}, we found $a=283230$ by assuming that the bandwidth at 80 m from the router was 5.5 Mbps, which matches the characteristics of a low-power edge network. We exclude $x, y \in (-1, 1)$ to satisfy the domain of Equation \ref{eq:bwdistance}, and to simplify the creation of our geometric graphs for our simulations. From a practical standpoint, this means that we assume that no devices will be within 1m of the router. 

We used the following configuration to test the algorithm:

\begin{enumerate}
    \item \textbf{Model}: MobileNetV2 \cite{mobilenetv2}, EfficientNetB1 \cite{efficientnet}, ResNet50, or InceptionResNetV2, chosen for their diversity in model topologies and memory sizes. Due to space constraints, we only show ResNet50 and InceptionResNetV2 in some results figures.
    \item \textbf{Number of Nodes}: 5, 10, 15, 20, or 50 randomly placed edge devices.
    \item \textbf{Number of Bandwidth Classes}: 2, 5, 8, 11, 14, 17, or 20 bandwidth classes, which provide granularity in how to classify the transfer sizes and edge bandwidths.
    \item \textbf{Node Memory Capacity}: 64, 128, 256, or 512 MB of RAM for a compute node.
\end{enumerate}

For each test, we used a different random communication graph generated using the procedure above. With each algorithm result, we then calculated the bottleneck latency according to Equation \ref{eq:latency}. The resulting bottleneck latency from each configuration of model, node capacity, number of nodes, and number of bandwidth classes was run 50 times and averaged. 

We compare the resulting bottleneck latency of our algorithm to that of the following two algorithms:

\begin{enumerate}
    \item \textbf{Random Algorithm}: Select a random node and a random partition that can be accommodated on that node.
    \item \textbf{Joint-Optimization Algorithm}: 
    Let $Q$ and $N$ represent the optimal set of partitions and optimal arrangement of nodes, respectively, chosen under this algorithm
    For each node $n$, do the following:
    \begin{enumerate}
        \item At each step choose the partition with the smallest transfer size that will fit within the node. Add this partition to the set of chosen partitions $p$.
        \item Starting at $n$, find the neighbor in the communication graph whose edge $e$ has the highest bandwidth, and add that to the path of chosen nodes $c$. Then, find the highest bandwidth edge from $e$, and so on.
        \item Compare the bottleneck latency found with $p$ and $c$ to the smallest bottleneck found with all nodes $n$ thus far, and update $Q$ and $N$ with $p$ and $c$ if the current bottleneck is smaller.
    \end{enumerate}
\end{enumerate}

For each of these algorithms, we used the same configuration and methodology as above to find the bottleneck latency. These algorithms don't use bandwidth classes, so we didn't need to include that as part of the configuration.

\section{Results}
\label{section:results}

\begin{figure}
\centerline{\includegraphics[scale=0.25]{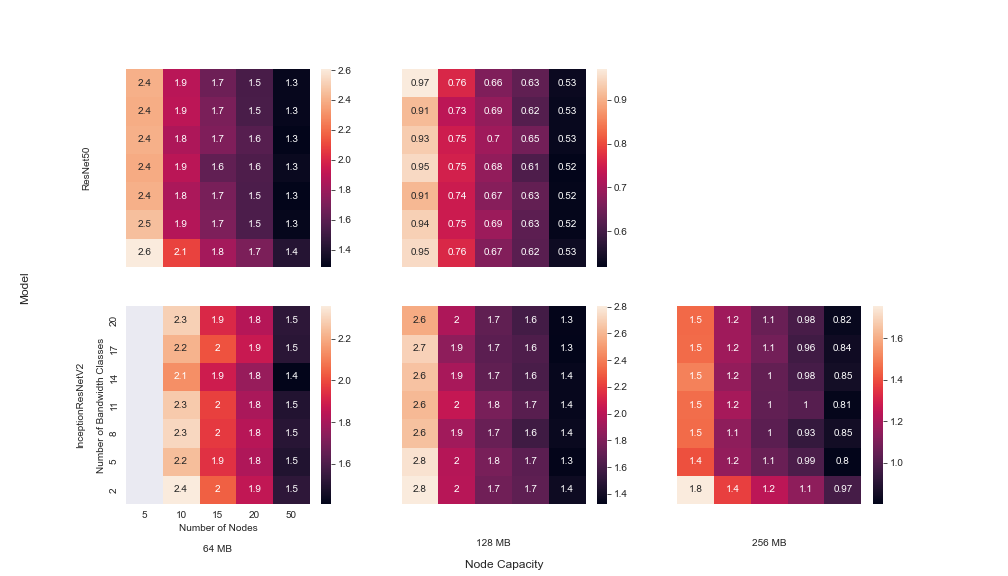}}
\caption{Color Map of Bottleneck Latency ($s$) based on Model, Node Capacity, Number of Nodes, and Number of Bandwidth Classes - Optimal Partitioning/Placement}
\label{fig:bottleneckplotsoptimal}
\end{figure}

Figure \ref{fig:bottleneckplotsoptimal} shows a color map of the resulting bottleneck latency based on the factors described in Section \ref{section:methods}. In Figure \ref{fig:bottleneckplotsoptimal}, the color map was only generated for the node capacities which were too small for the models to fit on a single device of that capacity. All models were able to fit on a single 512 MB device. The lack of bottleneck latency values for InceptionResNetV2 with 5 nodes and 64MB node capacity indicates that the model could not be partitioned with these physical constraints. For each model, the lowest bottleneck latency for a given node capacity comes from the combination of the most number of bandwidth classes and number of nodes. The lowest bottleneck latency comes with the highest node capacity. These results follow from the fact that a larger node and number of nodes allows the partitioning algorithm to have greater choice in selecting the smallest transfer sizes. Similarly, a high number of bandwidth classes allows the placement algorithm to better perform the $k$-path matching.
 
\begin{figure}
\centerline{\includegraphics[scale=0.25]{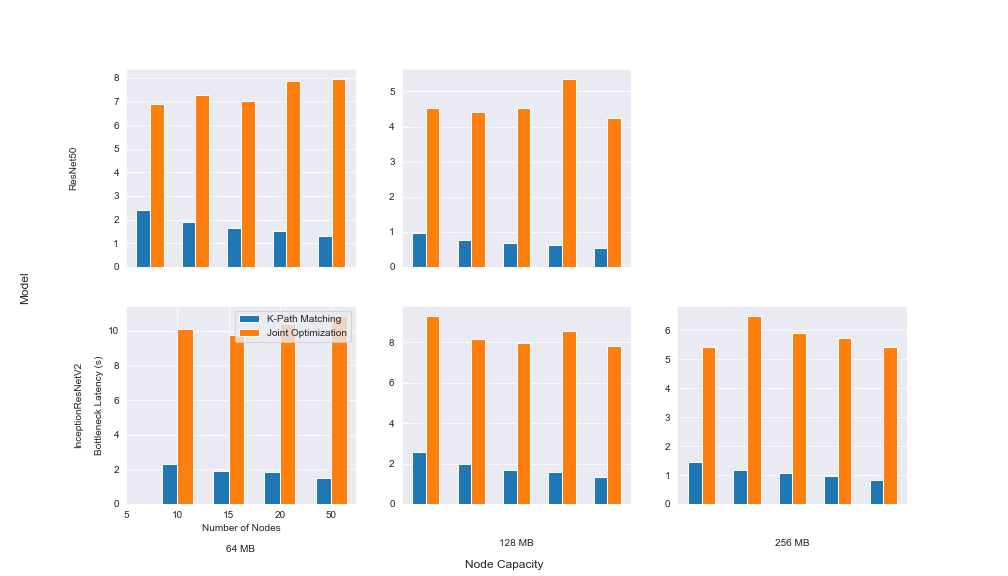}}
\caption{Comparison of Algorithm \ref{alg:kpathmatching} with Random Algorithm - based on Model, Node Capacity, Number of Nodes}
\label{fig:kpathvsrandom}
\end{figure}

In Figure \ref{fig:kpathvsrandom}, the optimal algorithm reduces bottleneck latency by $\approx 10$x on average for this selection of models. The difference is the smallest for ResNet50, with the optimal algorithm producing a $\approx 2$x lower bottleneck latency. The models with the greatest variance in transfer size will result in the largest difference in bottleneck latency between the optimal random algorithms. Overall, we see that the optimal algorithm produces a significant reduction in bottleneck latency compared to the random algorithm.

\begin{figure}
\centerline{\includegraphics[scale=0.24]{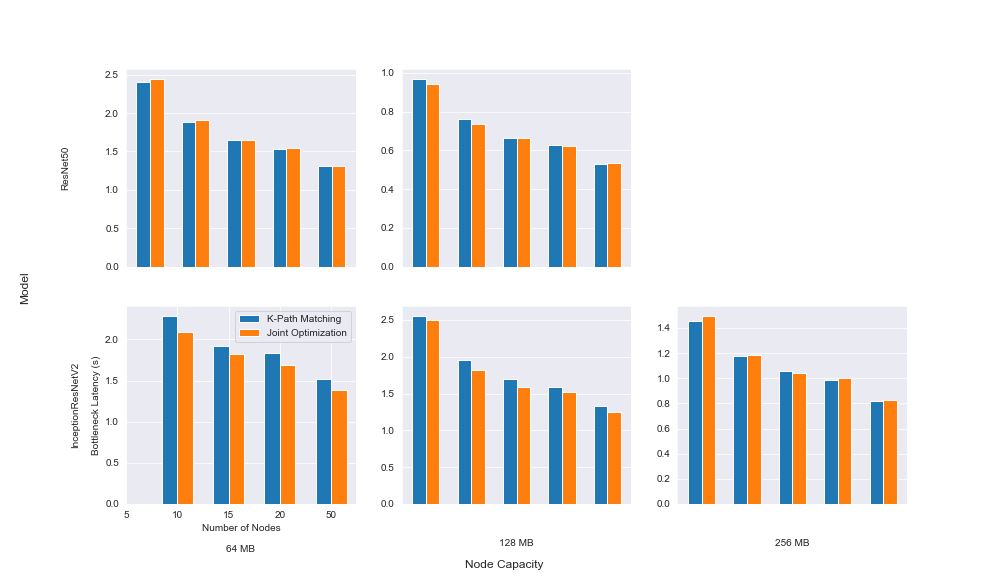}}
\caption{Comparison of Algorithm \ref{alg:kpathmatching} with Joint Optimization - based on Model, Node Capacity, Number of Nodes}
\label{fig:kpathvsjoint}
\end{figure}

In Figure \ref{fig:kpathvsjoint}, the joint optimization algorithm tends to perform better for a smaller number of nodes. Since each of these algorithms use the same optimal partitioning logic, we can only compare the models based on their differing placement logic. As the number of nodes increases, our $k$-path algorithm performs better. This makes sense, because the difference in the greedy strategy of the joint optimization algorithm and the matching strategy of our algorithm only becomes more apparent as the communication graph grows bigger and there are more options for node paths. In particular, for 50 nodes, our algorithm outperforms the joint optimization algorithm by 35\%. We hypothesize that this trend would continue for more complex models which have a greater number of candidate partition points and a greater variance in transfer size, necessitating the $k$-path matching strategy to minimize bottleneck latency. 

\begin{figure}
\centerline{\includegraphics[scale=0.4]{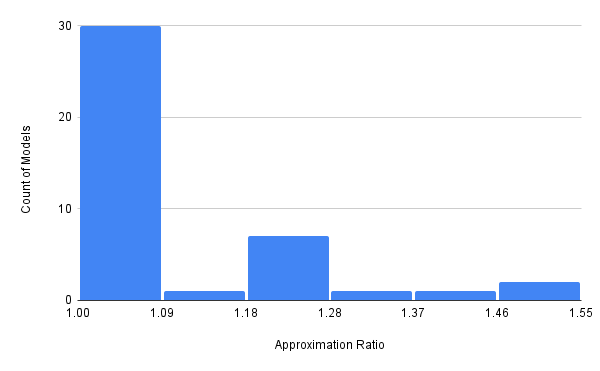}}
\caption{Histogram of Average Approximation Ratio for Keras Pretrained Models}
\label{fig:avgapproxratios}
\end{figure}

We then ran our algorithm 1000 times for the set of Keras pretrained models. We used a configuration of 50 nodes and 64 MB node memory capacity. For each trial, we then divided the resulting bottleneck latency by the optimal bottleneck latency as given by Theorem \ref{thm:latencybound}. We took the average of this ratio across all trials for each model. The results are presented in Figure \ref{fig:avgapproxratios}. We see that the bottleneck latencies for ~75\% of these models are within 9\% of the optimal solution. The average approximation ratio across all these models is $\approx 1.092$ or within 9.2\% of the optimal bottleneck latency.

\section{Conclusion}
We have presented a framework to partition and place a model across a set of resource-constrained edge devices, with the goal of maximizing inference throughput. Additionally, we show that for different models and node configurations, we can outperform a greedy joint-optimization algorithm. We further show empirically that our algorithm is within 9.2\% of the optimal bottleneck latency for the models we tested.

\subsection{Future Work}
With minor edits, we could extend our framework to work with geographically-distributed edge devices for a truly scalable edge inference solution.

With software changes, we could potentially run the average image model on a cluster of micro-controllers. We could use RiotOS \cite{riotos} without any containerization and perform optimizations to run with limited device memory. Some devices we could potentially take advantage of are the Raspberry Pi Pico \cite{rpi_pico} and Arduino Uno \cite{arduinouno}.

\section{Acknowledgements}
We would like to acknowledge the helpful input and pointers provided by Prof. Anil Vullikanti from the University of Virginia, particularly in directing us to the color-coding $k$-path algorithm. 

\bibliography{main}
\bibliographystyle{acm}
\end{document}